\documentclass[sigconf]{acmart}
\usepackage{caption}
\usepackage{subcaption}
\usepackage{multirow}
\usepackage{tabularx}
\AtBeginDocument{%
  \providecommand\BibTeX{{%
    \normalfont B\kern-0.5em{\scshape i\kern-0.25em b}\kern-0.8em\TeX}}}

\usepackage{array}
\newcolumntype{L}[1]{>{\raggedright\let\newline\\\arraybackslash\hspace{0pt}}m{#1}}
\newcolumntype{C}[1]{>{\centering\let\newline\\\arraybackslash\hspace{0pt}}m{#1}}
\newcolumntype{R}[1]{>{\raggedleft\let\newline\\\arraybackslash\hspace{0pt}}m{#1}}

\copyrightyear{2024}
\acmYear{2024}
\setcopyright{rightsretained}
\acmConference[RecSys '24]{18th ACM Conference on Recommender Systems}{October 14--18, 2024}{Bari, Italy}
\acmBooktitle{18th ACM Conference on Recommender Systems (RecSys '24), October 14--18, 2024, Bari, Italy}\acmDOI{10.1145/3640457.3691708}
\acmISBN{979-8-4007-0505-2/24/10}

\begin{document}

\title{Less is More: Towards Sustainability-Aware Persuasive Explanations in Recommender Systems}

\author{Thi Ngoc Trang Tran}
\affiliation{%
  \institution{Institute of Software Technology, Graz University of Technology}
  \city{Graz}
  \country{Austria}
}
\email{ttrang@ist.tugraz.at}
\orcid{0000-0002-3550-8352}

\author{Seda Polat-Erdeniz}
\affiliation{%
  \institution{Institute of Software Technology, Graz University of Technology}
  \city{Graz}
  \country{Austria}
}
\email{spolater@ist.tugraz.at}
\orcid{0009-0006-5106-3986}

\author{Alexander Felfernig}
\affiliation{%
  \institution{Institute of Software Technology, Graz University of Technology}
  \city{Graz}
  \country{Austria}
}
\email{alexander.felfernig@ist.tugraz.at}
\orcid{0000-0003-0108-3146}

\author{Sebastian Lubos}
\affiliation{%
  \institution{Institute of Software Technology, Graz University of Technology}
  \city{Graz}
  \country{Austria}
}
\email{slubos@ist.tugraz.at}
\orcid{0000-0002-5024-3786}

\author{Merfat El-Mansi}
\affiliation{%
  \institution{Institute of Software Technology, Graz University of Technology}
  \city{Graz}
  \country{Austria}
}
\email{merfat.el-mansi@student.tugraz.at}
\orcid{0009-0005-2695-4210}

\author{Viet-Man Le}
\affiliation{%
  \institution{Institute of Software Technology, Graz University of Technology}
  \city{Graz}
  \country{Austria}
}
\email{vietman.le@ist.tugraz.at}
\orcid{0000-0001-5778-975X}

\renewcommand{\shortauthors}{Tran, Polat-Erdeniz, Felfernig, Lubos, El-Mansi, and Le}

\renewcommand{\shorttitle}{Towards Sustainability-Aware Persuasive Explanations in Recommender Systems}

\begin{abstract}
Recommender systems play an important role in supporting the achievement of the United Nations sustainable development goals (SDGs). In recommender systems, explanations can support different goals, such as increasing a user's trust in a recommendation, persuading a user to purchase specific items, or increasing the understanding of the reasons behind a recommendation. In this paper, we discuss the concept of "sustainability-aware persuasive explanations" which we regard as a major concept to support the achievement of the mentioned SDGs. Such explanations are orthogonal to most existing explanation approaches since they focus on a "less is more" principle, which per se is not included in existing e-commerce platforms. Based on a user study in three item domains, we analyze the potential impacts of sustainability-aware persuasive explanations. The study results are promising regarding user acceptance and the potential impacts of such explanations.
\end{abstract}





\maketitle

\section{Introduction}\label{sec:introduction}
Recommender systems (RS) \cite{JannachDietmarRecommenderSystemsIntroduction} can contribute to the sustainable development goals (SDGs) defined by the United Nations\footnote{\url{https://sdgs.un.org/goals}} \cite{FelfernigetalSustainability2023}. Examples of these goals are \emph{responsible consumption and production}, \emph{good health and well-being}, and \emph{sustainable cities and communities}\footnote{For a more detailed overview, see Felfernig et al. \cite{FelfernigetalSustainability2023}.}. Related to the SDG of \emph{responsible consumption and production}, RS can be applied for recommending different options for supporting the reduction of energy consumption  \cite{Petkovetal2011,Spilloetal2023}. The SDG of \emph{good health and well-being} can be supported, for example, by the recommendation of healthy food items \cite{Tran2018JIIS}. Finally, RS can support the SDG of \emph{sustainable cities and communities}, for example, sustainable tourism RS supports the proposal of points of interest taking into account potential negative impacts on the environment, the local communities, and cultural heritage \cite{Banerjee2023,BaniketalSustainabilityTourism2023,Khanetal2021,Merinov2023}. 

In RS, explanations can be applied to achieve different goals \cite{ChangetalCrowdBasedExplanations2016,TintarevMasthoff2012,Koukietal2019,Mauroetal2023Justification}. Following the categorization of Tintarev et al. \cite{TintarevMasthoff2012}, examples of such goals are \emph{efficiency} (reducing the time that is needed to complete a choice task), \emph{persuasiveness} (to convince users to change his/her choice behavior), \emph{effectiveness} (to help users to make decisions of high quality), \emph{transparency} (making it transparent as to why specific items have been recommended), \emph{trust} (to increase a user's confidence in a recommender), \emph{scrutability} (providing ways to make the user profile adaptable), and \emph{satisfaction} (e.g., to increase the usability of the RS). These goals can be regarded as representative examples -- for related work we refer to \cite{FriedrichZanker2010,Guoetal2023,HerlockeretalExplainingCF2000,lubos2024llm,SharmaRay2016Overview,Tintarev2007,TintarevMasthoff2007,TintarevMasthoff2012,Yangetal2022,ZhangChen2020}.


In this paper, we show how to integrate sustainability aspects into persuasive explanations of RS. Related work on integrating persuasive principles into RS focuses on pushing the acceptance of items using persuasive messages  (see, e.g.,  \cite{AlslaityTran2021,GkikaLekakos2014,Yooetal2012}). For example, following the Cialdini persuasive principles \cite{Cialdini1993}, Alslaity and Tran \cite{AlslaityTran2021} analyze the impact of persuasion-aware item explanations on item selection behavior. Often, persuasive explanations appear more successful if an explanation refers to an item that is somehow ``compatible'' with the user's preferences \cite{GkikaLekakos2014}. Compared to the existing state-of-the-art of persuasive explanations in RS, we focus on the aspect of how to integrate sustainability aspects into explanations. Furthermore, we show how such explanations can be made ``persuasive'' by applying different persuasive principles to push users to make more sustainability-aware decisions.

To support SDGs, explanations have to focus on related argumentations. In car recommendation, an explanation could refer to the positive environmental effects of purchasing smaller cars. In video streaming, health-related consequences of extensive video consumption could be mentioned. In news recommendation, explanations could inform users about potential topic-wise biases with various social effects \cite{Milanoetal2020EthicalChallenges}. Following the \textit{``less is more''} principle, we introduce the concept of \textit{``sustainability-aware persuasive explanations''}. Potential impacts of such explanations could be, for example, fewer purchases of unnecessary items, the consumption of healthier food, and the purchase of items with an increased degree of sustainability (e.g., smaller cars). From the producer/company point of view, such explanations might be counterproductive, resulting in, for example, potentially decreasing turnovers. Such explanations contrast with mainstream recommender systems, which focus on increasing sales.

The contributions of this paper are as follows: (1) We introduce the concept of sustainability-aware persuasive explanations. (2) We provide a set of related example explanations in three item domains. (3) In a user study, we analyze the impact of such explanations and report corresponding study results.    

The remainder of this paper is organized as follows. In \textit{Section \ref{sec:sustainabilityawareexplanationsforrecommendersystems}}, we introduce Cialdini's persuasive principles \cite{Cialdini1993} as a basis for explanation generation. In this context, we provide explanations for related examples. In \textit{Section \ref{sec:userstudyresults}}, we describe the setting of our user study and discuss the study results. In \textit{Section \ref{sec:threatstovalidity}}, we discuss existing threats to validity. Finally, we conclude the paper with \textit{Section \ref{sec:conclusions}}.

\section{Persuasive Explanations}
\label{sec:sustainabilityawareexplanationsforrecommendersystems}
We now show how to apply Cialdini's persuasive principles (see \textit{Table \ref{tab:persuasiveprinciples}}) for sustainability-aware explanations. 


\begin{table}[ht]
\footnotesize
\centering \caption{Cialdini's persuasive principles \cite{Cialdini1993}. \vspace{-0.3cm}}
\begin{tabular}{|C{1.3cm}|C{6.5cm}|}  
\hline
  \textbf{principle}     &  \textbf{semantics}   \tabularnewline  \hline \hline
  reciprocity   &  feeling of an obligation to give something back \tabularnewline  \hline
  scarcity      &  reduced item availability increases preparedness to purchase  \tabularnewline  \hline   
  authority     &  experts have an increased influence over users \tabularnewline  \hline
  commitment    &  user prefer to be consistent with their articulated preferences \tabularnewline  \hline
  liking        &  users like other users who are similar to themselves \tabularnewline  \hline
  social proof  & users follow the opinions (of a representative set) of other users  \tabularnewline  \hline
\end{tabular}  \vspace{-0.4cm}
\label{tab:persuasiveprinciples} 
\end{table}


We follow the approach of Alslaity and Tran \cite{AlslaityTran2021}, who analyzed the impact of applying persuasive principles to achieve ``mainstream'' recommendation goals such as increasing item purchases. In contrast to \cite{AlslaityTran2021}, we focus on ``sustainability-aware persuasive explanations'' for triggering more sustainability-aware decisions. \textit{Tables \ref{tab:explanationsbooks}, \ref{tab:explanationshealthfood}}, and \textit{\ref{tab:explanationscar}} show examples of how these persuasive principles can be applied in recommendation settings.

\begin{table}[ht]
\footnotesize
\centering \caption{Sustainability-aware persuasive \emph{book} explanations with the SDG \textit{``responsible consumption and production''}. \vspace{-0.3cm}}
\begin{tabular}{|C{1.3cm}|p{6.6cm}|}  
\hline
  \textbf{principle}     &  \textbf{explanation}   \tabularnewline  \hline \hline
  reciprocity   &  The four books on topic X in your shopping cart appear to have a high content-wise overlap. Try to filter out at least one of those and you will receive a 25\% voucher for your next book purchase.\tabularnewline  \hline
  scarcity     &  Due to a content-wise overlap, reduce the number of books in your shopping cart and participate in a lottery to win a 50 EUR voucher for books you purchase online next time. Note that only two vouchers are still available. \tabularnewline  \hline   
  authority     &  Renowned book reviewers suggest that there is no need to purchase book A and book B (on topic X), since both have a 90\% content-wise overlap. The remaining 10\% are covered by article C which can be found here. \tabularnewline  \hline
  commitment    & Since you appear to be interested in responsible consumption, we recommend that you remove some books from your shopping cart since these appear to have a high content-wise overlap. It might make sense to analyze the related book reviews in more detail. \tabularnewline\hline
  liking        &  Customers with similar sustainability goals as you have decided to reduce their shopping cart, i.e., did not purchase all the popular books on topic X (also due to a high content-wise overlap). \tabularnewline  \hline
  social proof  & The majority of customers (with reading preferences similar to your own) interested in the topic X also decided to reduce their shopping cart, i.e., did not purchase all the existing popular books on the topic (also due to a high related content-wise overlap).  \tabularnewline  \hline
\end{tabular} \vspace{-0.3cm}
\label{tab:explanationsbooks} 
\end{table}

\section{Sustainability-Aware Explanations}\label{sec:userstudyresults}
\subsection{User Study Design}
We now present the design of our user study and provide details regarding \textit{participants' demographic information} and \textit{privacy aspects}.

\textit{\textbf{Item domains of user study}}. In our user study, we wanted to cover both low-involvement and high-involvement domains, which are often supported by different recommendation approaches \cite{FelfernigBurke2008}. As a low-involvement domain, we chose  \emph{books}, while \emph{cars} is considered a high-involvement item domain. The involvement of \emph{healthy food} as a third domain is ``located'' somewhere between books and cars. For book recommendation, basic approaches such as collaborative and content-based filtering are used quite often \cite{JannachDietmarRecommenderSystemsIntroduction}. For cars, knowledge-based RS are applied \cite{FelfernigBurke2008}. Our sustainability-aware explanations (\textit{Section \ref{sec:userstudyresults}}) are assumed to be applied directly to the recommended items, i.e., we do not assume a sustainability criteria-based item ranking.

\begin{table}
\footnotesize
\centering \caption{Sustainability-aware persuasive \emph{healthy food} explanations with the SDG \textit{``good health and well-being''}. \vspace{-0.3cm}}
\begin{tabular}{|C{1.3cm}|p{6.6cm}|}  
\hline
  \textbf{principle}     &  \textbf{explanation}   \tabularnewline  \hline \hline
  reciprocity   &  Decrease the overall family consumption of salty fish dishes by purchasing more fish dishes of type A and B. Through less salt consumption, you will experience many different advantages, for example, better sports performance and better well-being in general. \tabularnewline  \hline
  
  scarcity      &  Consuming unhealthy food (e.g., food with too much salt) can result in deteriorated health and well-being. The longer one waits to change his/her lifestyle, the higher the danger to experience the consequences of unhealthy food.  \tabularnewline  \hline
  
  authority     &  Based on reliable information from nutritionists, it appears that your family has been consuming an increased amount of salty fish dishes over the past four months. To ensure your good health, it is recommended that you decrease your consumption of these dishes. You may want to consider substituting them with alternative fish dishes, such as A and B, which have lower levels of salt. \tabularnewline  \hline
  
  commitment    & Since you appear to be interested in healthy living and well-being, we highly recommend to reduce your consumption of salty fish dishes we observed over the last four months and/or consume the ``less salinity'' fish dishes A and B. \tabularnewline  \hline
  
  liking        &  Customers also interested in the aspects of healthy living and well-being, switched their consumption from too high salty fish dishes to alternative fish dishes, such as A and B. \tabularnewline  \hline
  
  social proof  & The majority of customers who unintentionally consumed too many salty fish dishes over a longer time period, decided to switch to the alternative fish dishes A and B.  \tabularnewline  \hline
\end{tabular} \vspace{-0.6cm}
\label{tab:explanationshealthfood} 
\end{table}

\begin{table}
\footnotesize
\centering \caption{Sustainability-aware persuasive \textit{car} explanations with the SDG  \textit{``sustainable cities and communities''}. \vspace{-0.3cm}}
\begin{tabular}{|C{1.3cm}|p{6.6cm}|}  
\hline
  \textbf{principle}     &  \textbf{explanation}   \tabularnewline  \hline \hline
  reciprocity   &  Think about purchasing a smaller car, i.e., not an SUV. With this, you will contribute to sustainable cities of the future and also help to make life worth living for the upcoming generations. \tabularnewline  \hline
  
  scarcity      &  Purchasing an SUV is worth re-thinking, the longer we follow this line of mobility philosophy, the less opportunities we will have to achieve the goal of sustainable cities and communities.  \tabularnewline  \hline
  
  authority     &  Environmental scientists have emphasized recently that the overall share of SUVs significantly increases which results in an increased C02 emission and increasing parking infrastructure issues. Think about a smaller car since your goal is just to be mobile in the local environment.
  \tabularnewline  \hline
  
  commitment    & Since you declared your intention to contribute to a sustainable city of the future which includes the aspects of reduced CO2 emission and also efficient handling of parking infrastructures, you should think about purchasing a smaller car instead of an SUV. \tabularnewline  \hline
  
  liking        &  Customers interested in contributing to sustainable cities and communities finally decided to not purchase an SUV but to purchase a smaller car instead.  \tabularnewline  \hline

  social proof  &  Most customers interested in SUVs who just want to be mobile in the local environment often decided to switch from a larger to a smaller car model.  \tabularnewline  \hline
\end{tabular} \vspace{-0.8cm}
\label{tab:explanationscar} 
\end{table}


\textit{\textbf{Explanation preparation}}. The explanations designed to promote sustainability were developed by incorporating the persuasive principles and aligning these with SDGs for three specific item domains: \textit{books, healthy food,} and \textit{cars} (see \textit{Tables  \ref{tab:explanationsbooks}} to \textit{\ref{tab:explanationscar})}.

\textit{\textbf{Explanation distribution}}.
Using \textit{Google Forms}, we created three questionnaires (one per item domain) and distributed them to participants in a \textit{within-subject user study}. In each questionnaire, participants were asked to provide personal information such as \textit{age, gender,} and \textit{nationality}. Thereafter, a recommendation scenario was presented, explaining the recommendation context and the considered SDGs. Following this, \textit{six explanations} corresponding to the six persuasive principles (see \textit{Table \ref{tab:persuasiveprinciples}}) were shown to the participants. To minimize potential biases, the order of the explanations was \textit{randomized}, so that different participants received explanations in different orders.

Participants were then asked to review the explanations and rank each based on two dimensions: \textit{``persuasiveness''} (assessing how effectively the explanations convinced them to consume better and more sustainable choices) and \textit{``effectiveness''} (gauging how well the explanations enhanced their awareness of sustainable goals). Participants assigned a \textit{unique} ranking value from 1 (the lowest) to 6 (the highest) to each explanation. Besides, to gain further insights into the impacts of sustainability-aware explanations on user behavior changes, participants were asked to provide their opinions on two additional dimensions: \textit{``importance''} and \textit{``influence''}. The \textit{importance} dimension measures the significance of including sustainability aspects in product recommendations, while the \textit{influence} dimension assesses how sustainability-aware explanations affected participants' online purchasing behavior. Participants used a five-level rating scale (\textit{1: unimportance/no influence} and \textit{5: very high importance/very high influence}) to express their opinions.

\textit{\textbf{Participants and demographic information}}. We gathered complete responses from \textit{158 participants}, consisting of \textit{118} males (74.68\%) and \textit{40} females. The distribution of participants across each domain was as follows: books (\textit{53}), healthy food (\textit{55}), and cars(\textit{50}). Slight variations occurred due to some participants not completing their responses, leading to exclusion from the final analysis. Most participants were students in our university course aged 18-24 (136 participants, 86.08\%), 25-34 (21 participants, 13.29\%), and only one participant aged 35-44 (0.63\%). A high number of participants were from Austria (114 participants, 72.15\%), followed by Germany (5 participants, 3.16\%), and other European countries (39 participants, 24.68\%). For study participation, we only required proficiency in English, as the study was conducted in English and did not require specialized domain knowledge.

\textit{\textbf{Privacy aspects}}. Participants were informed about the privacy policy before joining the user study ("study-only" data usage, no sharing with third parties, and deletion after study completion). 





\subsection{Data Analysis}

\textit{\textbf{Methods}.} Our primary focus is to compare the \textit{``persuasiveness''} and \textit{``effectiveness''} of the proposed sustainability-aware persuasive explanations in promoting healthier and more sustainable choices. To achieve this, we collected participants' rankings of the explanations based on these two dimensions across three specified item domains. Thereafter, we categorized the rankings into three groups: \textit{low ranking} (values 1 and 2), \textit{average ranking} (values 3 and 4), and \textit{high ranking} (values 5 and 6). We then compared the results at two levels: \textit{within each domain (level 1)} and \textit{across different domains (level 2)}. For \textit{level 1}, in each item domain, we performed cross-tabulations for each dimension to display the number of participants who assigned each ranking value to the explanations. For \textit{level 2}, we conducted a \textit{one-way ANOVA test} ($\alpha=0.05$) for each explanation to analyze (significant) differences across the item domains.

Another focus of our study is to investigate the \textit{``importance''} and \textit{``influence''} of the explanations on product recommendations and users' purchasing behaviors. To gain deeper insights, for each dimension, we collected \textit{three} evaluation sets corresponding to three item domains. The evaluations in each set share the same characteristics such as \textit{ordinal variables} $\in [1..5]$, \textit{independent} of each other (the evaluations of the dimension in one domain did not rely upon those of other domains), and \textit{not normally distributed} (\textit{Shapiro-Wilk} tests, $\alpha = 0.05$, $p-values < \alpha$). For this reason, we used the \textit{Kruskal-Wallis test} ($\alpha = 0.05$) to analyze our data. For each dimension, we conducted a Kruskal-Wallis test to determine if there were significant differences across the three domains. We also examined the mean ranks generated by this test to identify the domain with the highest importance/influence (i.e., the higher the mean rank, the greater the importance/influence). If the Kruskal-Wallis test indicated a significant result, we conducted follow-up pairwise tests (\textit{Mann-Whitney U tests}) to identify significant differences between pairs of item domains. Since three Mann-Whitney U tests are needed for three item domains, we applied a \textit{Bonferroni adjustment} to avoid \textit{Type I errors} \cite{Pallant2007}. This adjustment modified the significance level to $\alpha' = 0.05/3 = 0.0167$.

\begin{figure*}
    \centering 
    \includegraphics[width=0.75\textwidth]{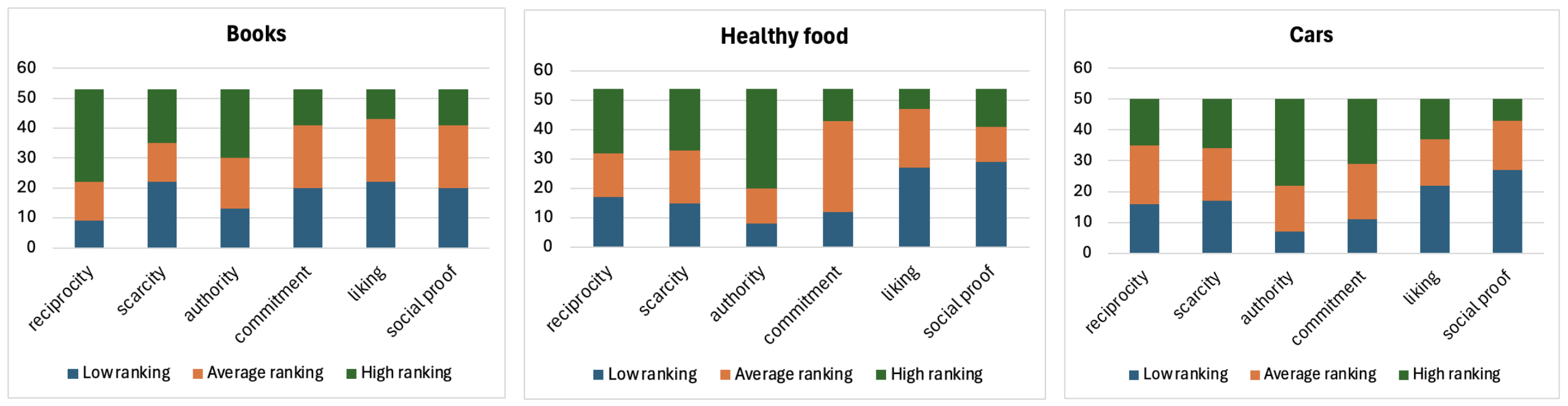} \vspace{-0.3cm}
    \caption{The persuasiveness of the sustainability-aware persuasive explanations across three item domains. The y-axes show the number of participants who ranked the explanations. \vspace{-0.3cm}}
    \label{fig:persuasiveness_level1}
\end{figure*}

\begin{figure*}
    \centering
    \includegraphics[width=0.75\textwidth]{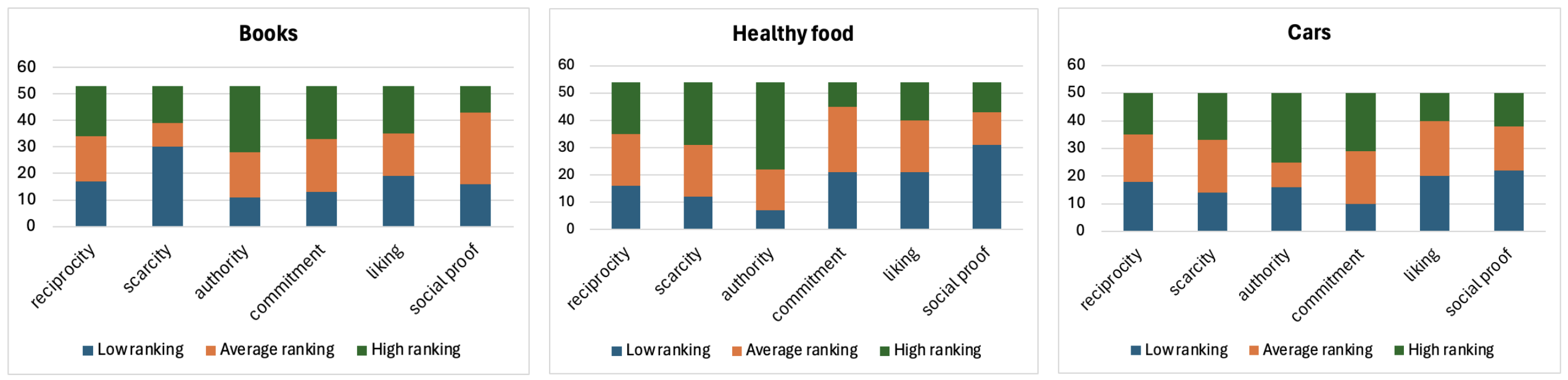} \vspace{-0.3cm}
    \caption{The effectiveness of sustainability-aware persuasive explanations across three item domains. The y-axes show the number of participants who ranked the explanations.} \vspace{-0.4cm}
    \label{fig:effectiveness_level1}
\end{figure*}

\textit{\textbf{Results}.} \textbf{\textit{Persuasiveness and effectiveness}}. Regarding \textit{persuasiveness}, the level-1 analysis revealed consistent trends in the healthy food and car domains. Explanations based on the \textit{``authority''} principle received high rankings, while those utilizing the \textit{``social proof''} principle were ranked the lowest. However, this pattern differed in the book domain, where explanations using the \textit{``reciprocity''} principle received the highest rankings, and those employing the \textit{``scarcity''} and \textit{``liking''} principles were ranked the lowest. Besides, explanations based on the \textit{``commitment''} principle generally received average rankings in all domains (see \textit{Figure \ref{fig:persuasiveness_level1}}). Concerning \textit{effectiveness}, explanations based on the \textit{``authority''} principle again received the highest rankings in all domains. However, the average and lowest rankings varied across domains. In the book domain, the lowest rankings were often assigned to the \textit{``scarcity''} explanations, whereas in the other two domains, the lowest rankings were given to the \textit{``social proof''} explanations. Similarly, average rankings were assigned to different explanations: the \textit{``social proof''} explanation in the book domain, the \textit{``commitment''} explanation in the healthy food domain, and the \textit{``liking''} explanation in the car domain (see \textit{Figure \ref{fig:effectiveness_level1}}). In the level-2 analysis, the ANOVA tests revealed significant results only for a few explanations, such as the \textit{``reciprocity''} explanations ($p_{reciprocity} = 0.011$) for the \textit{persuasiveness} dimension, and the \textit{``scarcity''} and \textit{``commitment''} explanations ($p_{\text{scarcity}}=0.001$ and $p_{\text{commitment}}=0.015$) for the \textit{effectiveness} dimension. However, based on the mean ranking values from these tests, the \textit{``authority''} explanations consistently received the highest rankings for both dimensions across the three item domains (see \textit{Tables \ref{tab:persuasiveness_level2} \& \ref{tab:effectiveness_level2}}).

\textbf{\textit{Importance and influence}}. The Kruskal-Wallis test for the \textit{``importance''} dimension showed a significant result ($p = 0.05$), indicating significant differences in participant evaluations across the item domains. However, the follow-up pairwise tests (Mann-Whitney U tests, $\alpha' = 0.0167$) did not reveal significant differences between any two domains. The inspection of the mean ranks indicated that sustainability aspects in product recommendations were considered more important in the healthy food and car domains than in the book domain (see \textit{Table \ref{tab:mean_rank}}). Similarly, the Kruskal-Wallis test for the \textit{``influence''} dimension did not yield significant results. Nonetheless, the mean rank analysis showed a similar trend to the \textit{``importance''} dimension, with sustainability aspects having a greater impact on average- and high-involvement domains (see \textit{Table \ref{tab:mean_rank}}).

\begin{table}[ht]
\footnotesize
\centering \caption{The mean ranking of the explanations regarding \textbf{\textit{persuasiveness}} across the item domains. \vspace{-0.3cm}}
\begin{tabular}{|c|c|c|c|c|c|p{0.5cm}|}  
\hline
    &  reciprocity & scarcity & authority & commitment & liking & social proof \tabularnewline  \hline \hline
  books  & \textbf{4.40}	& 3.19 & 4.11 &	3.15 &	3.00 & 3.15  \tabularnewline  \hline  
  food  & 3.72 & 3.67 &	\textbf{4.52} & 3,44 & 2.85 & 2.80  \tabularnewline  \hline 
  
  cars  &  3.38 & 3.44 & \textbf{4.44} &	3.88 &	3.08 & 2.78 \tabularnewline  \hline 
  
\end{tabular} \vspace{-0.6cm}
\label{tab:persuasiveness_level2} 
\end{table}

\begin{table}[ht]
\footnotesize
\centering \caption{The mean ranking of the explanations regarding \textbf{\textit{effectiveness}} across the item domains. \vspace{-0.3cm}}
\begin{tabular}{|c|c|c|c|c|c|p{0.5cm}|}  
\hline
    &  reciprocity & scarcity & authority & commitment & liking & social proof \tabularnewline  \hline \hline
  books  & 3.64	& 2.74 & \textbf{4.13} &	3.74 & 3.49 & 3.26 \tabularnewline  \hline  
  
  food  & 3.63 & 3.94 &	\textbf{4.57} & 3.06 & 3.24 & 2.63  \tabularnewline  \hline 
  
  cars  &  3.32	& 3.64 & \textbf{3.90} &	3.86 & 3.10 & 3.16 \tabularnewline  \hline 
  
\end{tabular}  \vspace{-0.6cm}
\label{tab:effectiveness_level2} 
\end{table}

\begin{table}[ht]
\footnotesize
\centering \caption{Mean ranks created by Kruskal-Wallis tests for the dimensions \textit{``importance''} and \textit{``influence''} in the item domains. \vspace{-0.5cm}}
\begin{tabular}{|c|c|c|}  
\hline
   domain &  importance & influence \tabularnewline  \hline \hline
  books  & 67.14 & 72.43 \tabularnewline  \hline  
  
  food  & \textbf{85.53} & \textbf{87.03} \tabularnewline  \hline 
  
  cars  & \textbf{84.52} & \textbf{77.29} \tabularnewline  \hline 
  
\end{tabular} \vspace{-0.2cm}
\label{tab:mean_rank} 
\end{table}

\textit{\textbf{Discussion}.} Our data analysis revealed item domain-specific differences in participants' rankings for  \textit{``persuasiveness''} and \textit{''effectiveness''}. However, a common tendency across average- and high-involvement item domains has emerged. Explanations based on the \textit{``authority''} principle consistently ranked as the most persuasive and effective. Conversely, explanations utilizing the \textit{``commitment''} principle received average rankings, while those based on \textit{``social proof''} received the lowest rankings. The high rankings of \textit{``authority''} principle-based explanations can be attributed to users' trust in expert opinions and authoritative sources, particularly in average- and high-involvement domains such as healthy food and cars, where professional endorsements carry substantial weight. Conversely, users' skepticism of peer influence led to the lowest rankings for \textit{``social proof''} explanations. 

Regarding the \textit{``importance''} and \textit{``influence''} of importing sustainability aspects in product recommendations and users' purchasing behavior, users also placed greater importance on sustainability in average- and high-involvement item domains. This could be explained by higher stakes and broader societal impact associated with decisions in these domains. For example, healthier food choices contribute to personal well-being and to community health. Similarly, choosing sustainable transportation options benefits both the individual and the environment. Conversely, the lower perceived impact of sustainability in the book domain may stem from the lower environmental footprint of book production and distribution, leading consumers to perceive sustainability as less relevant.

\section{Threats to Validity \& Future Work}\label{sec:threatstovalidity}
To ensure item domain diversity, we chose one example, a low-involvement item domain (books), one high-involvement item domain (cars), and one ``in-between'' item domain (healthy food). To increase the result's generalizability, we will analyze further domains with a more diversified set of participants. Furthermore, our user study is based on a synthesized setting. For future work, we plan user studies with real-world items. Finally, we are aware that we conducted a user study with a selected set of SDGs (\emph{responsible consumption and production}, \emph{good health and well-being}, and \emph{sustainable cities and communities}). We regard further user studies covering all $17$ SDGs as an important goal of our future work.

\section{Conclusions}\label{sec:conclusions}
To support sustainability-aware decision-making in recommendation contexts, we have introduced \emph{sustainability-aware persuasive explanations}. Using persuasive principles, we have generated example explanations for three item domains and evaluated how users perceive such explanations. We have analyzed which persuasive principles are perceived as the most impactful ones by users and if users would appreciate the provision of such explanations.


\bibliographystyle{ACM-Reference-Format}
\bibliography{bibliography}

\end{document}